\documentclass[a4paper, amsfonts, amssymb, amsmath, reprint, showkeys, nofootinbib, twoside]{revtex4-1}
\usepackage[english]{babel}
\usepackage[utf8]{inputenc}
\usepackage[colorinlistoftodos, color=green!40, prependcaption]{todonotes}
\usepackage[pdftex, pdftitle={Article}, pdfauthor={Author}]{hyperref} 
\usepackage{float}
\usepackage{stackengine}
\begin{document}
\title{
 Entropic alignment of topologically modified ring polymers in cylindrical confinement}

\author{Sanjay Bhandarkar$^1$, Debarshi Mitra$^1$, J\"urgen Horbach$^2$ and Apratim Chatterji$^1$}
    \email[Correspondence email address: ]{apratim@iiserpune.ac.in}
    \affiliation{ $^1$ Department of Physics, IISER-Pune, Pune, Dr. Homi  Bhabha Road, India-411008}
    \affiliation{$^2$ Institut f\"ur Theoretische Physik II-Soft Matter, Heinrich-Heine-Universit\"at D\"usseldorf,
    Universit\"atsstrasse 1, D-40225 Düsseldorf, Germany}
\date{16 November, 2025}

%
\begin{abstract}
%

Under high cylindrical confinement, segments of ring polymers can be localized along the long axis 
of the cylinder by introducing internal loops within the ring polymer. The emergent organization of 
the polymer segments occurs because of the entropic repulsion between internal loops [Phys.Rev.E, 
{\bf 106}, 054502 (2022)]. These principles were used to identify the underlying mechanism of bacterial 
chromosome organization [Soft Matter {\bf 18}, 5615-5631 (2022)]. 
Here, we outline functional principles associated with entropic interactions, leading to specific orientations 
of the ring polymers relative to their neighbors in the cylindrical confinement. We achieve this by modifying 
the ring polymer  topology by creating internal loops of two different sizes within the polymer, and thus create an asymmetry. 
This allows us to strategically manipulate polymer topology  such that 
segments of a polymer face  certain other segments of a neighboring polymer. The polymers therefore 
behave as if they are subjected to an `effective' entropic interaction reminiscent of interactions 
between Ising spins. But this emergent spatial and orientational organization is not enthalpy-driven. 
We consider a bead spring model of flexible polymers with only repulsive excluded
volume interactions between the monomers. 
The polymers entropically repel  each other and occupy  different halves of the cylinder, and moreover, the adjacent polymers preferentially re-orient themselves along the axis of the cylinder. 
We further substantiate our observations by free energy calculations. 
To the best of our knowledge, this is the first study of the emergence of effective orientational 
interactions by harnessing entropic interactions in flexible polymers.  
The principles elucidated here could be relevant to understand the interactions between different sized loops within a large chromosome.  
\end{abstract}

\maketitle 

\section{Introduction}
The bacterial cell of {\em E.coli} uses entropic mechanisms to segregate its two daughter chromosomes to two halves of the cell \cite{Jun2010,Jun2007,dna1}. The cell then divides to produce two daughter cells, each with one copy of the chromosome. The bacterial cell does not have the protein machinery called mitotic spindles, which is present in the eukaryotic cell of higher organisms. This protein machinery pulls the replicated daughter chromosomes to segregate them to two halves of the eukaryotic cell. In bacterial cells such as {\em E.coli}, the process of replication of the mother chromosome to two daughter chromosomes and the segregation of the two daughter DNA-polymers occur simultaneously. The entropic segregation of ring polymers under cylindrical confinement has been proposed as one of the key mechanisms governing the spatial segregation and organization of two newly formed daughter chromosomes in sphero-cylindrical {\em E. coli} cells \cite{Jun2012,Badrinarayanan2015,Woldringh_r_2024}. Bacterial chromosomes are ring polymers, and like most bacterial cells {\em E.coli} has a single chromosome. 

Under cylindrical confinement, ring polymers tend to spontaneously segregate for entropical reasons. If two overlapping ring polymers are considered in a cylinder with only excluded volume interactions between the monomers, the two ring polymers may segregate to two halves of the cylinder \cite{Jun2006,Jun2010} to maximize their conformational entropy. Thus, the polymers can explore a higher number of conformations (microstates) if they are segregated along the long axis of the cylinder, compared to a state where they overlap \cite{Arnold2007, Frenkel2007, Jun2007, Junprl2008, Jung2009}. This happens in an infinite cylinder as well as for a finite cylinder of suitable aspect ratio with closed ends \cite{ha1_2010, ha2_2012,ha3_2012}. For the entropy-driven segregation of (ring) polymers, it is crucial to have a high degree of confinement i.e., the diameter of the cylinder must be smaller than the radius of gyration $R_g$ of the unconfined ring polymer. The same phenomenon can also happen for a linear polymer, but the entropic forces of segregation, as well as tendency to remain demixed are much higher for a ring polymer \cite{dna2}.  

For the bacterial cell {\em E.coli}, the underlying physical mechanism of the segregation of daughter chromosomes was not well understood for a long time. An important step forward has been the work of Jun and Wright \cite{Jun2010} who proposed that the leading mechanism for the segregation of chromosomes could be revealed by studying the entropic segregation of ring polymers in cylindrical confinement. Furthermore, as suggested later \cite{Harju2024,chase,dna2}, modification of the topology of the ring polymer due to linker proteins or SMC proteins that extrude loops may enhance entropic forces of segregation for two polymers in a cylinder \cite{Harju2024,chase,dna2}. It is known that particular proteins are necessary for the successful segregation of daughter chromosomes in the {\em E.coli} cell and it has been suggested that these proteins play a role in creating loops \cite{Sherratt2020,Harju2024,chase}. In our past studies, we have modified the ring polymer topology by introducing cross-links between specific monomers of the ring polymer. The introduction of these cross-links creates loops of polymer segments. The cross-links are created such that the ring polymer contains multiple internal ring-like loops on the chain contour. Using our specifically designed polymer topologies, we were also able to entropically localize certain segments or monomers of the polymer at specific sections along the long axis of the cylinder. Some topologies are better than others to keep polymers de-mixed (prevent partial overlap of monomers) near the center of the cylinder. A detailed investigation of the enhanced segregation and emergent localization properties of ring polymers as a consequence of topological modifications, has been detailed in Ref.~\cite{dna2}. The results from this work have been used to describe the underlying mechanism of the spatio-temporal chromosome organization in the {\em E.coli} cell \cite{dna1,dna_fast}.

From our previous work \cite{dna1,dna_fast}, we have evidence that the replicating bacterial chromosome adopts a particular modified ring polymer topology, as the cell goes through its life cycle. Additional transient topological modifications can occur at smaller length scales along the contour by linker or SMC proteins, apart from the global topological modifications that we consider in this work. We model the replication of topologically modified (ToMo) mother chromosome, and obtain the segregation of the two ToMo daughter chromosomes in our simulations, even when the cylindrical cell elongates to twice its length before division. The aspect ratio of the confining cylinder that we choose for our initial investigations \cite{dna2} also correspond to the aspect ratio of sphero-cylindrical {\em E. coli} cell with two fully replicated daughter DNA-polymers. Thereby, we can explain the mechanism underlying the localization of chromosome loci as seen in FISH experiments with the {\em E.~coli} cell. 

While there has been persistent interest in knotted polymers \cite{muthu1,muthu2,Kardar1,kardar2,Virnau1,knot_review}, interest has peaked in the last few years in the study of topologically modified polymers \cite{Brackley2013,Marenduzzo2006,Bonato2021,Tubiana2024,Mondal2017,Polson2018,dieterloop}. This is because of their relevance in understanding the basic properties and functions of chromosomes in both eukaryotic and prokaryotic cells \cite{Conforto2024,Brahmachari24956,Sherratt2020,Schiessel2023,Hofmann2019,Polovnikov2023,Haddad2017,DAsaro2024,DiStefano2021,Mithun2,Kadam2023,Agarwal2018,Agarwal2019,Agarwal2019_2}. Furthermore, as our previous studies establish, emergent properties as a consequence of modifying the topology of a pair of ring-polymers, is useful to obtain a mechanistic understanding of the organization and segregation of daughter chromosomes \cite{dna1,dna2,dna_fast}. Hence, it is of interest that we explore further consequences and emergent physics of modifying the topology of confined polymers, which may be later useful to understand observed  experimental phenomena related to chromosomes. This study is a step in that direction. The proposition that entropy leads to the organization of polymers is particularly significant from a theoretical point of view because the idea lends itself naturally to a variety of contexts related to synthetic (non-replicating) polymers, without the necessity of assuming specific interactions between monomers. 

The key new question that we address in this manuscript is about an entropy-based mechanism by which a system of flexible polymers can be arranged in a certain manner. Here, the idea is that the entropic interactions between polymers can be ``designed'' via the internal topology of the polymers. Flexible polymers chains are considered as intrinsically disordered objects,  but we can break symmetry in the internal conformational phase space of a single ring-polymer by strategically introducing cross-links between two monomers separated along the contour to have internal loops of different sizes. Thereby, we can create an asymmetric polymer architecture (topology), such that one section is notionally called the head (H) and the other end is referred to as the tail (T). In addition, we can confine these polymers in a long cylinder, where radial confinement is considerably  more than the confinement along the long axis. Thereafter, the organization of two (or more) polymers along the long axis can be controlled in a manner that is reminiscent to that of spins interacting {\em via} the Ising  Hamiltonian, if one is able to define a head (``up spin'') and tail (``down spin'') of a polymer. If the orientational  preference of the two polymers with respect to each other is driven by entropy, then we cannot have a $T=0$ energy minimum configuration. In this work, we establish that we can indeed topologically modify polymers suitably to have emergent effective interactions between polymer sections in the confined geometry of a cylinder, such that polymers have higher probability to choose particular configurations with respect to each other.  While there is already a huge body of work on the statics and dynamics of linear and ring polymers under confinement \cite{Shin2014,Shin2015,elena1,elena2,Halverson2014,Halverson2011_1,Halverson2011_2,Smrek2021,Smrek2015,Zhou2019,Vettorel2009,Schram2019,Rosa2019,Jost2018,Rosa2008,Rosa2010,Ravi2022, Narros2010,blobtheory,Chubak2018, Chubak2021} we focus on emergent properties on further introduction of internal loops within a ring polymer confined in a cylinder.

\begin{figure*}
\includegraphics[scale = 0.85]{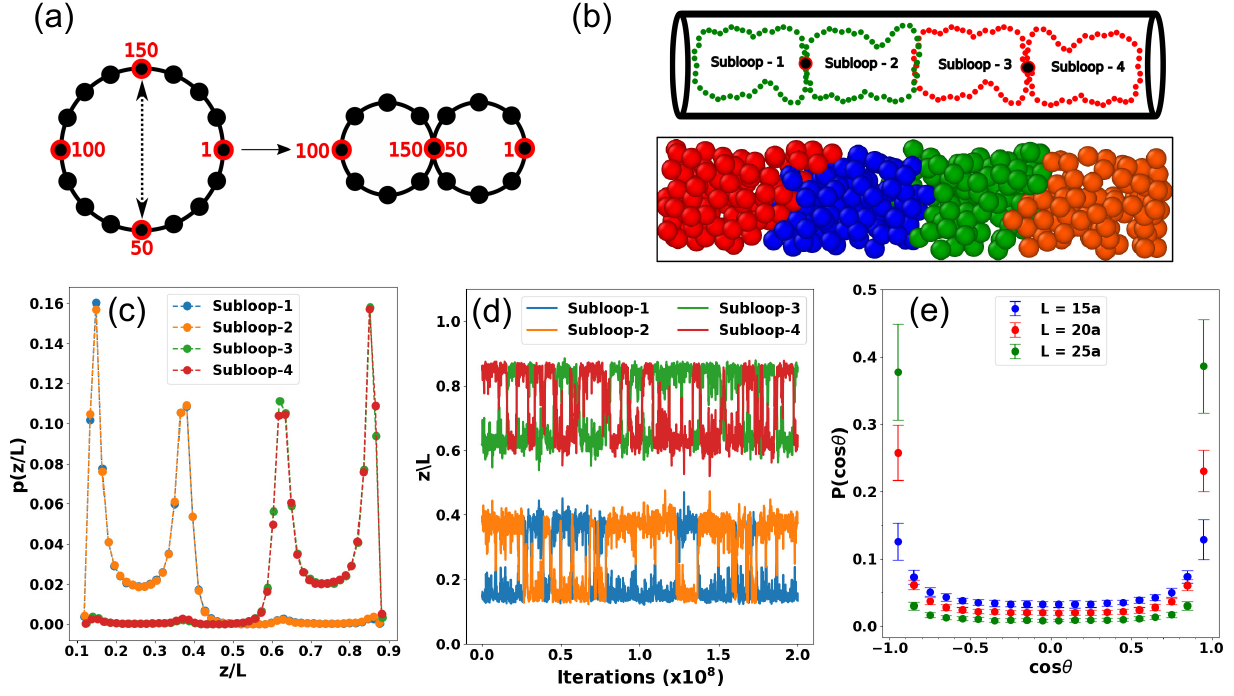}
\caption{(a) A schematic of a ring polymer that is topologically modified to a `rotated-8' with 200 monomers in each ring, (b) Different subloops of the two polymers along the long axis of the cylinder. The snapshot from the simulations shows the different subloops in different colors. (c) Probability density distribution $p(z)$ of the center of mass (COMs) of the different loops, denoted by subloop-1, 2, 3, 4. The coordinate $z$ is along the long axis of the cylinder. (d) Position of COM of each subloop as a function of number of iterations (simulation time).  The data re-iterates that the loop-COMs stay well separated along $z$, though the subloops of a polymer interchange positions along $z$. (e) Statistical average of the dot product of the two vectors joining the COMs of the subloops of the two polymers as well as plot the probability density of $\langle \cos(\theta) \rangle$. \label{fig1}}
\end{figure*}

In the upcoming sections, we begin with a model Section and thereafter proceed to study the properties of a pair simple modified ring polymers, with just one cross-link (CL) introduced in each polymer. Here, both the loops are of equal size, and this sets up the basic understanding of how the internal loops of polymers arrange themselves along the long axis of the cylinder as they interact with each other, and the loops of other polymers, as well as with the confining walls. Thereafter, we consider ring polymers with $2$ internal CLs, thereby creating three internal loops and an asymmetry within a single polymer. Then, we increase the asymmetry within a single polymer by introducing more CLs, and investigate the emergence of entropy-driven orientational organization within a pair of polymers, such that certain segments of a polymer prefer to be in contact with certain other segments of the neighbouring polymer. In Section IV, we computationally calculate free energy differences between the preferred and other configurations in a pair of polymers with various topological modifications. In Section V, we analyze the contact maps of such topologically modified segregated polymers in a cylinder, with the expectation that the contact maps will be able to appreciate more complex scenarios with topologically modified polymers in future. We finally end with a discussion in Section-VII.

\section{Model}
We perform Langevin dynamics simulation of a flexible, bead-spring model of polymer chains. This model considers a chain of spherical beads, where neighboring beads are connected by mass-less harmonic springs of stiffness $\kappa$ and equilibrium length $a$. The harmonic spring potential is given by $V_{spring}$ = $\kappa(r-a)^2$, where $\kappa$ = 100$k_BT$/$a^2$. We consider $a$ to be the unit of length in our simulation, and energies are measured in units of thermal energy $k_BT$. Furthermore, the excluded volume interactions between the monomers are modelled by the Weeks-Chandler-Andersen (WCA) potential:
\begin{equation}    
     \centering
    V_{WCA} = 4 {\epsilon} \left[\left(\frac{\sigma}{r}\right)^{12}- \left(\frac{\sigma}{r}\right)^{6} \right] + {\epsilon}_0, \hskip0.15cm \forall r < r_c=2^{1/6}{\sigma},
\end{equation}
where $\epsilon = k_BT$, $\sigma = 0.8a$, and $\epsilon_0$ is added to ensure the purely repulsive potential goes to zero smoothly at $r_c$.

We consider two or more identical topologically modified ring polymers confined within a cylinder for our studies. The interactions between the monomers and the walls of the cylinder are also modeled by a WCA potential. We vary the length $L$ and the diameter $D_c =2R_c$ of the cylinder.  We always choose the $D_c$ to be significantly smaller than the radius of gyration $R_g$ of an unconfined ring polymer. We work with polymer chains which have $N=200$ monomers, unless specified otherwise. Two polymers with $200$ monomers each in a cylinder with $D_c=5a$ and $L=25a$ results in monomer volume fraction of $\approx 0.2$. The $R_g$ for a $200$ monomer (unconfined) ring polymer in good solvent conditions with excluded volume interactions is calculated as $\approx \sqrt{f} \times R_g(L) = \sqrt{f} \times (a N^{\nu}/\sqrt{(1+2\nu)(2+2\nu)}) \approx 6.7a$, where $\nu=0.6$ is the Flory exponent of a self avoiding walk (SAW) polymer chain. The factor $f=0.55$ is the ratio $R_{g, \mathrm{ring}}^2/R_{g, \mathrm{lin}}^2$ of the radii of gyration $R_{g, \mathrm{ring}}$ and $R_{g, \mathrm{lin}}$ for a polymer ring and a linear chain, respectively. We introduce cross-links (CLs) between specific monomers of a ring polymer to design different topological modifications. The cross-linking is achieved by introducing harmonic springs between the specific monomers, maintaining the same bond length and $\kappa$, as specified above. Thereby, internal loops within the ring polymers are obtained. The details of the topological modifications are given in Sec.~III. The radius of gyration of the topologically modified ring polymers with $N=200$ monomers is different from that of a polymer ring with the same number of monomers. We use, however, the same cylinder size as mentioned above to confine the modified polymers to maintain the number density of monomers. When comparing data from distinct computations of polymer properties with different topology or $N$, we prefer to keep the monomer volume fraction fixed, instead of keeping the ratio $D_c/R_g$ constant, because we are aiming at maintaining monomer volume fractions close to what is seen {\em in-vivo} for {\em E. coli} chromosomes.

We use Langevin dynamics as implemented in LAMMPS \cite{LAMMPS} to realize our model system. We use a time step of $\delta t = 0.01 \tau_0$, where $\tau_0 = \sqrt{m a^2/k_BT}$ is the unit of time in our simulations. We choose the mass of the monomers to be $m = 1$. A Langevin thermostat maintains the system at a fixed temperature $T$ with a damping constant $\Gamma = m/\tau = \tau_0^{-1}$. 

We initialize the rings while ensuring they are already segregated along the cylinder's axis and are not concatenated. To achieve this, we initialize the monomers along a compact circle whose diameter is smaller than that of the confining cylinder. However, this leads to a significant overlap of monomers, and consequently initial configuration have a high energy. Hence, we first perform $10^4$ Monte-Carlo steps (MCS) to decrease the overlap and thereby reduce the overlap energies. Thereafter, we allow our system to equilibrate for $10^8$ iterations in Langevin simulations, and only then begin to collect data for statistical analysis in a subsequent production run over $10^8$ iterations. The results presented below are obtained from an average over $40$ statistically independent runs. 

\begin{figure}[ht!]
\includegraphics[scale = 0.9]{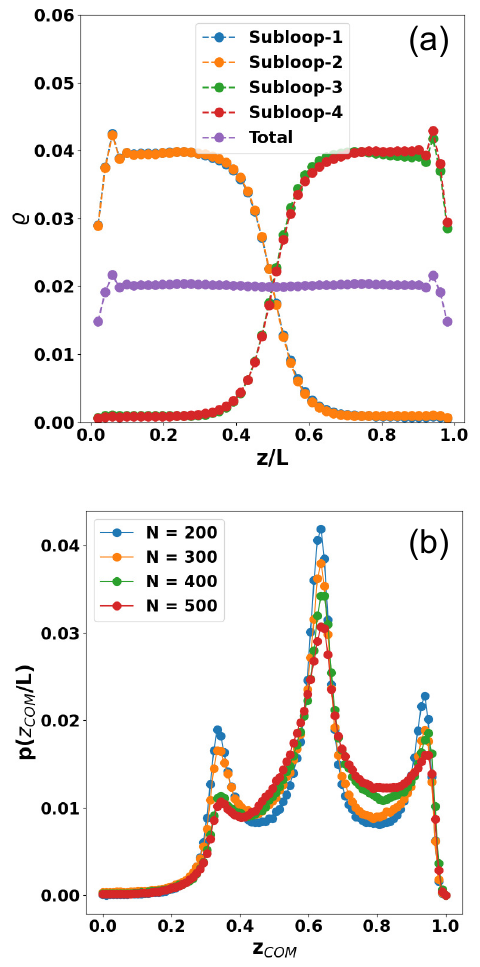}
\caption{(a) Mean density of monomers belonging to different subloops of two polymers of `rotated 8' architecture as a function of $z$. Each polymer has $N=200$ monomers. Data is normalized by the number of monomers ($100$) in each loop. The legend `Total' refers to the mean monomer density of the two polymers, normalized by the total number of monomers in the cylinder ($400$). (b) Probability distribution of the distance between the COMs of subLoop-1 and subLoop-3 with varying number of monomers $N$ of the polymer. The cylinder diameters, chosen for `rotated 8' polymers with $N=200$, 300, 400, and 500 monomers, are $D_c= 5a$, $5.9a$, $6.48a$, and $7a$, respectively. The length of the cylinder, $L$, is chosen such that $D_c/L=5$. \label{fig2}}
\end{figure}
To ensure that the system has reached equilibrium, we analyze the $z$-coordinate of the COM of the different segments of each polymer to confirm that the internal loops of each polymer flip and interchange their positions multiple times along the length of the cylinder over the course of the equilibration run. This process ensures that the polymers have relaxed to equilibrium before we collect data to calculate statistical quantities. We have also explicitly checked that $10^6 \tau_0$ is well above the longest conformational relaxation times of the ring polymer in free space. For ease of communication, monomers from $1$ to $50$ to $100$ (and monomers from $101$ to $150$ to $200$ to $1$) are associated with the right (or left) arm of the ring.

\section{Results}
We begin with two ring polymers of size $N=200$ each, confined in a cylinder of length $L=25a$ and diameter $D_c=5a$. This results in an aspect ratio of 1:5 and a volume fraction of $\phi$ = $0.218$, where
\begin{equation}
    \phi = \frac{4 N {(\sigma/2)}^3}{3 R_c^2 L}
\end{equation}
The diameter of the cylinder is chosen such that $D_c$ is smaller than the equilibrium size $2 R_g \approx 14a$ of an unconfined ring polymer.  

{\em Rotated-8 topology:} We add a topological modification to the polymers by adding a cross-link (CL) between the 50th and 150th monomers of each polymer, see the schematic of Fig.~\ref{fig1}a. This CL brings the diametrically opposite monomers close to one another, effectively `pinching' the ring polymer to form two smaller subloops. In the following, we refer to this topologically modified polymer as the `rotated $8$' topology.  A schematic representation of the organization of subloops is shown in Fig.~\ref{fig1}b, and a snapshot from our simulations confirms that the polymer loops remain primarily demixed. 

We calculate the center-of-mass (COM) positions of each subloop and plot the probability density distribution functions, $p(z/L)$, of the COM positions along the $z$-axis in Fig.~\ref{fig1}c. For each subloop, the distribution $p(z)$ shows four distinct peaks. The occurrence of these peak is a consequence of the entropic repulsion between the subloops such that the different subloops occupy different segments of the cylinder. Thus, different segments of the polymers are well separated and localized to different quarters of the cylinder. However, the subloops can flip and exchange positions. This can be inferred from the overlapping distributions for subloop-1 and subloop-2 for one of the polymers (Fig.~\ref{fig1}c). Similarly, the distributions for subloop-3 and subloop-4 for the other polymer also overlap. This can also be clearly seen if we follow the COM of each subloop as a function of time (Fig.~\ref{fig1}d). The polymers remain demixed with the flipping time of COMs being relatively small. This implies that a relatively small free energy barrier has to be crossed for the subloops of a polymer to exchange positions (see below). There are also rare occasions where the polymers inter-change positions over the length of a run, as indicated by the probability density distributions (PDD) for subloop-1 and subloop-2 (and subloop-3 and subloop-4) that also have non-zero values on the other half of the cylinder (Fig.~\ref{fig1}b). The probabilities of the loops which are closer to the two ends (poles) of the cylinder have higher probabilities due to wall effects.

To further probe the localization of loops along the cylinder axis, we define a vector between the COM of subloop-1 and subloop-2 (and correspondingly, subloop-3 and subloop-4) for each polymer (cf.~Fig.~\ref{fig1}b). Via the dot product of the two vectors, we compute $\cos(\theta)$, with $\theta$ the angle between the two vectors. In Fig.~\ref{fig1}e, we show the probability distribution of $\cos(\theta)$, $P(\cos(\theta))$. This distribution exhibits pronounced maxima at $\cos(\theta) = \pm 1$, indicating that the subloops organize such that the vectors are mostly parallel or anti-parallel to each other.  

An important point to note is the difference in peak heights in the COM distributions $p(z/L)$ (see Fig.~\ref{fig1}c). The two peaks closer to the center have a smaller amplitude compared to the loops located at the ends of the cylinder. Furthermore, the width of the peaks are different. Moreover, the overlap of $p(z/L)$ for different subloops within the same polymer is larger than the overlap between subloops of different polymers: we observe that $p(z/L =0.5) \approx 0$. To better understand this feature, we plot in Fig.~\ref{fig2} the monomer density of each subloop. At the poles of the cylinder, the monomer density for each loop shows a small peak which is due to hard repulsion of the monomers by the walls. The repulsive interaction with the walls is also responsible for the sharp drop of $p(z/L)$ near the poles. Near the middle of the cylinder, the monomers from one polymer overlap with the monomers of the other polymer. Furthermore, there can be an overlap between monomers belonging to different loops of the same polymer, as can be inferred from the distributions in Fig.~\ref{fig1}c. 

\begin{figure*}[ht!]
\includegraphics[scale = 0.8]{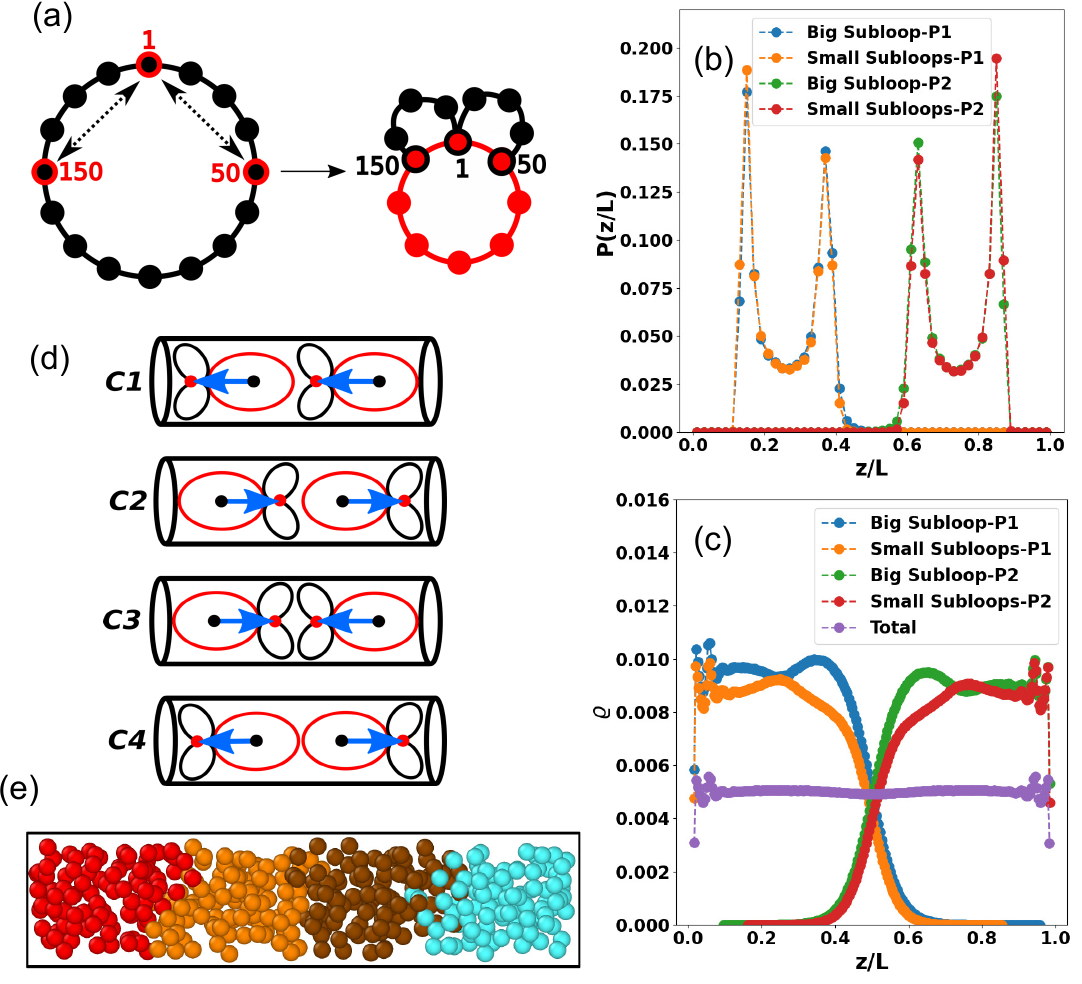} 
\caption {(a) Schematic of a ring polymer with $N=200$. The tip of the arrows show the position of the monomers on the contour which are cross-linked such that an Arc-1-2 topology is obtained. The polymer has a big subloop (red beads) and two small subloops (black beads). (b) Probability density distribution $p(z)$ of the center of mass (COM) of the monomers of different loops, from a pair of polymers, referred to as $P1$ and $P2$, respectively. The confining cylinder has length $L=25a$ and diameter $D_c=5a$. (c) Mean density of monomers from different loops along the $z$ axis. (d) Schematic of idealized configurations, classified as C1, C2, C3, and C4, that are predominantly attained by a pair of polymers having the Arc-1-2 topology. The big subloop and the two Small subloops are shown as red and black lines, respectively. The blue vector joins the COM of the bigger loop to the crossing point of the two smaller subloops for each polymer. (e) Representative snapshot of two Arc-1-2 polymers from the simulation. Here, the monomers of smaller and bigger loops are shown in different colors. \label{fig3}}    
\end{figure*}

Thus, the difference in the peak heights of $p(z/L)$ (Fig.\ref{fig1}b) is due to the fact that a subloop at the center overlaps and faces soft entropic repulsion from two subloops on either side. However, a subloop present towards the poles of the cylinder faces repulsion from a subloop on one side and a hard repulsive wall on the other side. As a result, the COM of the subloops, located near the centers, is spread over a wider range of $z$ as compared to the subloop at the corner. This consequently leads to a difference in the peak heights. 

In Fig.~\ref{fig2}b, we also show for different lengths $N$ of polymers the probability distribution $p(z_{COM})$ for the COMs of subloop-1 \& subloop-3 (from two different polymers) to be at a distance $z_{COM}$. The distribution calculated for the distance between the COMs of subloop-2 and subloop-4 is identical to the data shown for subloop-1 and subloop-3. For each value of $N$, the distribution has three peaks. The left-most peak at $z/L \approx 0.35$ corresponds to the case when both the subloop-1 and subloop-3 occupy positions near the center of the cylinder and thus the loops can overlap. The peak at $z_{COM}/L \approx 0.9$ corresponds to a configuration where subloop-1 and subloop-3 are both near the poles of the cylinder. The second peak in the middle corresponds to the case where one of the two subloops is near the poles and the other one is located near the center in the other half of the cylinder. This configuration can arise when either subloop-1 or subloop-3 is near the can be near the poles while the other loop is far away from the poles. As a consequence, the peak near $z_{COM}/L \approx 0.6$ has nearly twice the amplitude compared to the two other peaks. We see that that loops remain well segregated along the long axis and an overlap of loops is entropically penalized. Longer polymers with $N=300$, 400, 500 monomers and correspondingly bigger loops lead to broader distributions $p(z_{COM}/L)$, indicating more overlap between monomers of different loops. Note also that with increasing $N$ the ratio of the peak heights increases (e.g.~for $N=500$ the amplitude of the central peak is nearly a factor of three larger than those of the side peaks).

{\em Arc-1-2 topology:} Having established the localization of subloops with two `rotated-8' polymers, we now consider systems with further modified polymer topology. As illustrated in Fig.~\ref{fig3}a, we again begin with a pair of ring polymers, each with $N=200$ monomers. Thereafter, we cross-link monomers $1$ \& $50$ and monomers $1$ \& $150$ to create a polymer with modified Arc-1-2 topology. In this topology, we consider a system with one big subloop of $100$ monomers and two small subloops with $50$ monomers each. In the the nomenclature of the Arc-1-2 architecture, the first (second) number denotes the number of large (small) loops. Just to set the convention, later we will also refer to this (and other) architectures as Arc-1-2 [100-50], where the two numbers in square brackets indicate the number of monomers in the big loop and the two small loops, respectively. However, we will use this convention only when we are comparing data for polymers with different number of monomers of the same architecture.

To investigate polymer-loop organization with Arc-1-2 architectures in cylindrical confinement, we calculate the COM positions of the $100$ monomers of the big subloop as well as the COM of the two small subloops (combined), as the two small subloops together also have a total of $100$ monomers. Then, we define vectors from the COM of the big subloop to the COM of the two small subloops. We also plot the probability distribution of the COM of the big subloop and COM of the two smaller subloops along the $z$-axis (Fig.~\ref{fig3}b). As seen in the investigations with a pair of `rotated-8' polymers, we observe that (i) the two polymers are well segregated along $z$; (ii) the COM of different loops show four distinct peaks in $p(z)$, indicating that the positions of loops for a particular configuration are also well segregated in space. 

\begin{figure}[ht!]
\includegraphics[scale = 0.55]{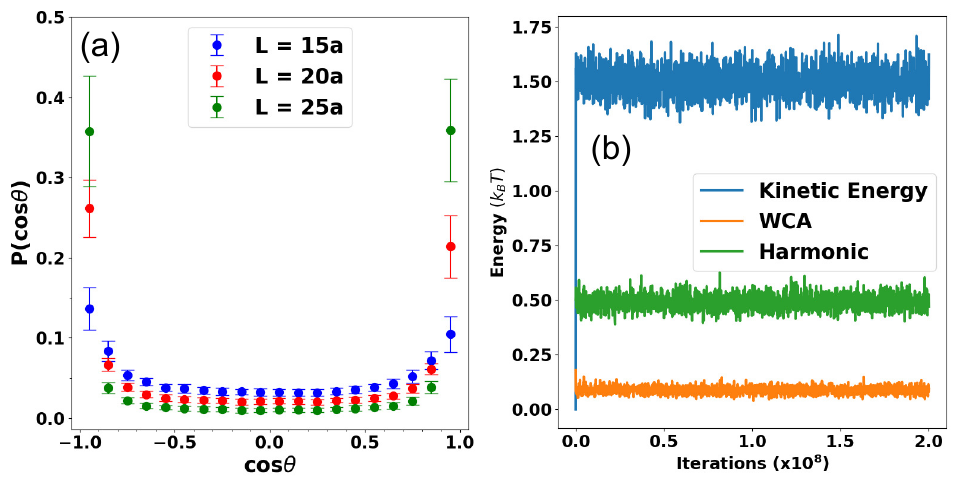}
\caption{We define the angle $\theta$ between the two vectors, one for each polymer, which joins the COM of the monomers of the big loop to the COM of the monomers of the two small loops (combined) of the same polymer. (a) Probability distribution $P(\cos \theta)$. The quantity $\cos(\theta)$ is the dot product of the two vectors for a particular configuration. The small asymmetry in $P(\cos(\theta))$ becomes more prominent as we decrease from $L=25a$ to  $L=20a$ and $L=15a$. (b) Relative contributions (per monomer) of the spring potential between neighboring monomers along the chain contour, the excluded volume interactions, and the kinetic energy as a function of time for a pair of polymer in a cylinder of length $L=25a$ and diameter $D_c=5a$. The contribution of excluded volume interactions is minimal. \label{fig4}}
\end{figure}

Figure \ref{fig3}c shows the monomer density of each of the subloops for Arc-1-2 topology of polymers confined in a cylinder of $L=25a$. In the legend, {\em 'total'} refers to the average of the four monomer densities for the two big subloops and four small subloops, normalized by the total number of monomers in the two polymers, i.e., $2N$. There is a distinct difference in the monomer density distribution in each half of the cylinder along the long axis, indicating that the monomers of two smaller loops spread out when they occupy the center of the cylinder. The monomers of the the two polymers only exhibit an overlap near $z/L\sim 0.5$ (Fig.~\ref{fig3}c).  The difference in the distribution of monomer density for polymers P1 and P2 along the cylinder axis could be attributed to an unequal number of flips of polymers P1 and P2, even when averaged over $40$ independent  runs. If one observes carefully the position distribution of the $p(z/L)$ of the COM of the loops, the height of the peaks of $p(z/L)$ in panel (b) is different for the small loops and the big loops.

Now that we have introduced an anisotropy in one of the subloops of the polymer, we can classify the conformations attained by a system of two Arc-1-2 polymers into the four configurations shown in Fig.~\ref{fig3}d. Any other configuration that involves significant mixing of the internal loops of the two polymers, will be placed under `Others'. In such cases, the COM of a subloop from polymer P2 could lie between the COMs of the subloops of polymer P1. As shown in Fig.~\ref{fig3}d, in configurations $C1$ \& $C2$ the vectors joining the COM of the sub-loops for the two polymers are parallel, while in configurations $C3$ \& $C4$, the vectors are anti-parallel to one another. Representative snapshot of two Arc-1-2 polymers from our simulations shown in Fig.~\ref{fig3}e reconfirms that the polymer loops have minimal overlap along the long axis $\hat{z}$.

\begin{figure}[ht!]
\includegraphics[scale = 0.55]{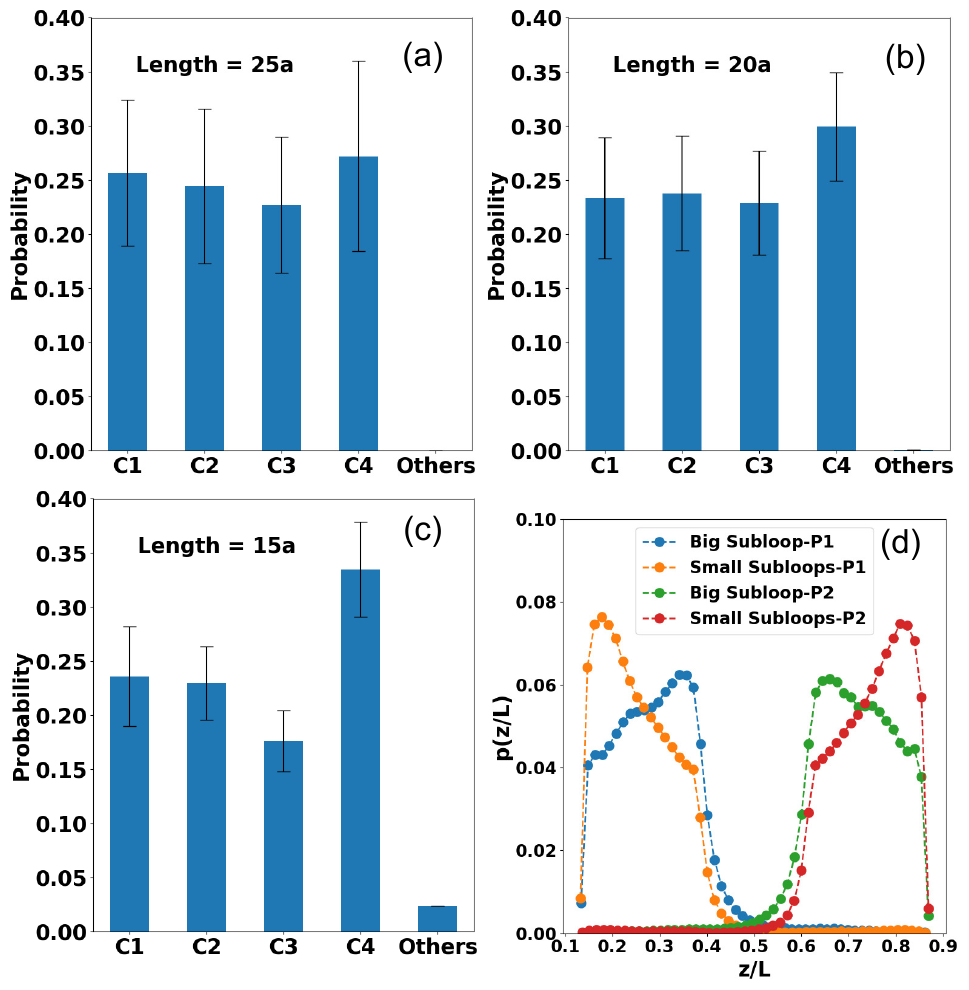}
\caption{The histograms show the probabilities of different configurations in panels (a), (b), (c), respectively for cylinders of lengths $L=25a, L=20a$, $L=15a$ and in all case for a fixed diameter $D_c=5a$. Two Arc-1-2 polymers confined in a cylinder prefers to entropically organize themselves in the anti-parallel C4 configuration. The error bars denote a standard deviation (SD), as obtained from a calculation over 40 independent runs. In panel (d), we show the probability density distributions of COM of the big subloops and the two small subloops of two polymers P1, P2, confined in a cylinder of length $L=15a$. \label{fig5}}
\end{figure}

In Fig.~\ref{fig4}a, we plot the probability distribution of $\cos(\theta)$, which is a measure of how the subloops are positioned along the $\hat{z}$ axis. We calculate $\cos(\theta)$ by taking the dot product of the two vectors joining the COMs of subloops (as shown schematically by the blue vectors in Fig.~\ref{fig3}d) for each snapshot and plot the distribution of the value of $\cos(\theta)$, obtained for each microstate in Fig.~\ref{fig4}(a). The distribution now shows an asymmetry in the probabilities of $\cos(\theta) = 1$ and $\cos(\theta) = -1$. To investigate whether an increased confinement enhances this effect, we repeat the calculation for cylinder lengths of $20a$ and $15a$, keeping the diameter  unchanged. Obviously, with decreasing $L$ the asymmetry becomes more pronounced, indicating higher preference for anti-parallel configurations, i.e., C3 \& C4. In Fig.~\ref{fig4}b, we also display the interaction energy between the subloops in units of $k_BT$. We see that the repulsive WCA interaction between monomers is relatively small compared to the other contributions. Thus, the loops remain the loops remain segregated primarily to maximize their conformational entropy. The peaks in $P(\cos(\theta))$ give an indication of the `orientation' of a polymer with respect to the other.

Next,  we plot the probabilities of finding the polymers in one of the configurations C1, C2, C3, and C4. There is a difference in probabilities of configurations C3 and C4 (Fig.~\ref{fig5}a). We can differentiate between C3 and C4 from the directions of the two blue vectors shown in Fig.~\ref{fig3}d. In Fig.~\ref{fig5}(b) and (c), the probabilities of the four configurations are shown for different values of $L$. For smaller cylinders,the difference in the probabilities of C3 and C4 gets magnified such that the occurrence of the C4 configuration has a higher probability corresponding to a decrease in the probability of C3. Since C1 and C2 are equivalent in each case, their probabilities are equal. We also note that  the probability of `Others' also increases as the propensity of mixing of the loops of the polymers increases as we decrease the aspect ratio of the cylinder. We do not consider cylinders of smaller lengths as it would lead to significant overlap (mixing) of the polymers segments in such denser systems. To develop a better understanding of the behavior of polymer segments, we plot the probability distribution of the COM positions of the subloops along the $z$-axis for two polymers in a cylinder of length $L = 15a$ (Fig.~\ref{fig5}d). Here, we observe higher probabilities for bigger subloops to occupy the middle of the cylinder. Below, we present calculations of free energy differences where we compare free energy of configurations where big loops overlap vis a vis scenarios when the smaller loops overlap. But one can intuitively guess that there will be enhanced entropic cost when four $50-$monomer loops face each other at the center of the cylinder in C3. It is entropically favourable for bigger `softer' loops to overlap and explore a large number of configurations, even as they occupy the center of cylinder \cite{Narros2013,blobtheory}.
 
\begin{figure}[ht!]
\includegraphics[scale = 0.55]{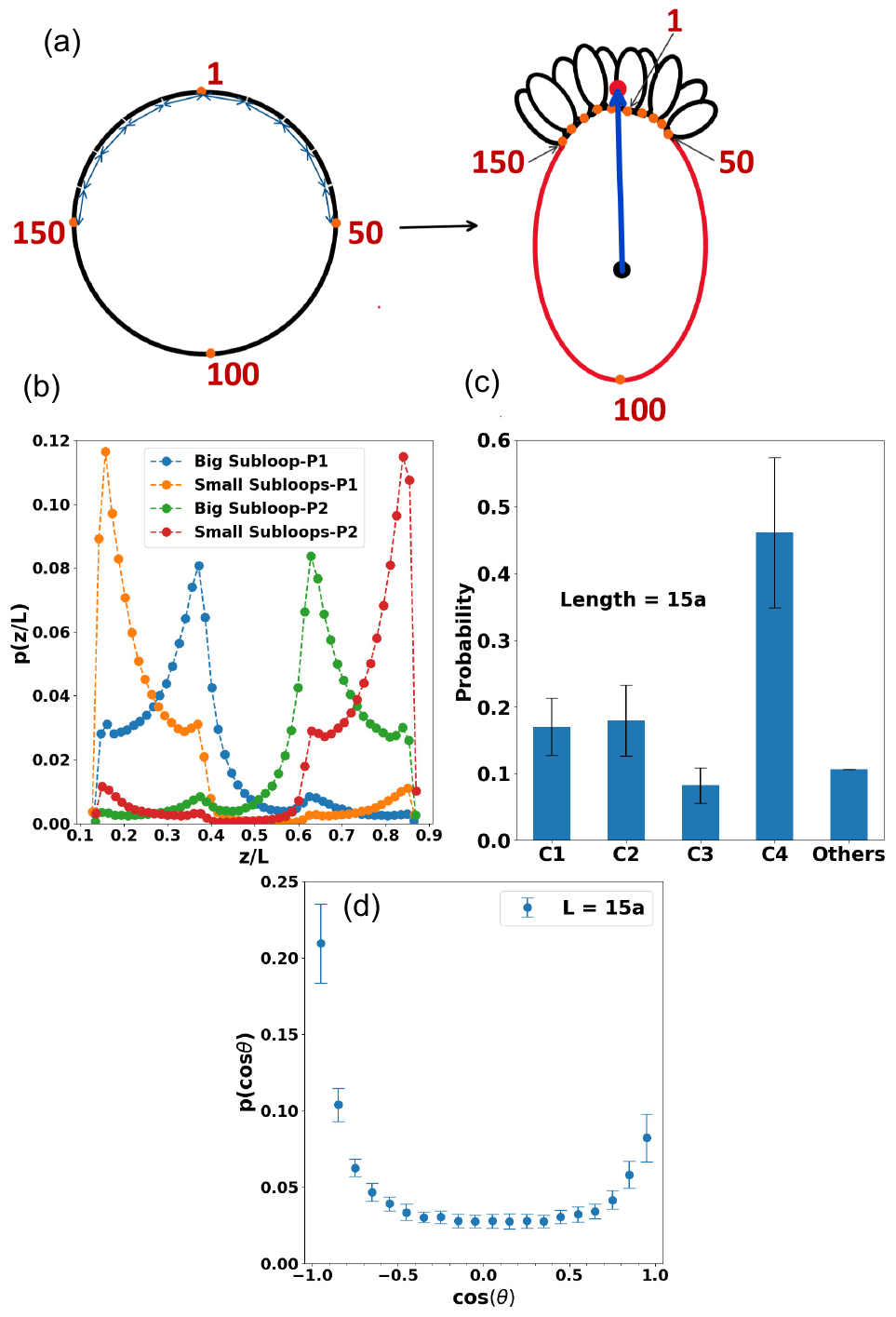}
\caption {(a) Schematic of the Arc-1-10 polymer architecture. The arrow tips represent the position of monomers along the contour of a $N=200$ ring polymer, which are cross-linked to produce the $10$ small loops (black ellipses), each containing $10$ monomers. (b) Probability density distribution of the COM of the big subloops, as well as the monomers of the ten small subloops of two polymers P1 and P2, confined in a cylinder. (c) Probability for the occurrence of configurations C1-C4 for $L=15a$ and $D_c=5a$. (d) Probability distribution of the dot-product of the vectors joining the COM of the big and small subloops. \label{fig6}}
\end{figure}

{\em Arc-1-10 topology}: Now we create the Arc-1-10 topology by adding 10 CLs in each ring polymer. To create this topology, we join the monomers $1$ and $10$ by a spring, i.e.~create CL $(1-10)$ and also $(11-20)$, $(21-30)$, $(31-40)$, $(41-50)$. On the other arm, we cross-link $(151-160)$, $(161-170)$, $(171-180)$, $(181-190)$, $(191-200)$. This leads to the formation of one big subloop of $100$ monomers and ten smaller subloops of size $10$ each. Note that monomer $200$ and $1$ are adjacent along the contour of the ring polymer, and thus monomers $51$ to $150$ form the big loop. A schematic of this geometry is provided in Fig.~\ref{fig6}a. Thus, each on the left and the right arm there are $5$ small loops. The big loop covers half of the left as well as half of the right arm.

The probability density distribution of the COM positions for the big subloop and all the monomers of the ten small subloops (combined) is plotted in Fig.~\ref{fig6}b. Again, there are four peaks, now with marked contrast between the probability density distributions of the COM of the big subloop and that of small subloops. We observe that the big subloops have higher probability to occupy the center of cylinder as compared to the distribution for the cluster of small subloops. This can be contrasted with the probability distributions shown in Fig.~\ref{fig1}b and in Fig.~\ref{fig3}b for the architectures considered previously. This preference for the big subloops to occupy the center of the cylinder consequently forces the set of small loops to preferentially occupy positions towards the poles of the cylinder. It is important to note that we do encounter configurations where the polymer P1 and P2 have interchanged positions along the long axis. The observation of a non-zero probabilities for the COM of the subloops of the first polymer (P1) in the right half of the cylinder (and vice versa) in Fig.~\ref{fig6}b is indicative of this conclusion. Interestingly, we observe that the big subloops tend to remain at the center even after exchange of polymer positions along the long axis.

To quantify the preference of C4 configuration as compared to others, we classify each configuration and display probabilities of different configurations C1, C2, C3, and C4 in Fig.~\ref{fig6}c. We see a clear preference for the C4 configuration, with the C4 being about five times more probable than C3 and more than twice as probable than the C1 and C2 configurations individually. We then plot the distribution of $\cos(\theta)$ in Fig.~\ref{fig6}d, calculated by the dot product of the vectors joining the center of the subloops. The distribution of $\cos(\theta)$ again shows a stronger asymmetry of the probabilities, and a clear preference for the C4 and C3 configuration with $\cos(\theta) = -1$ being three times more probable than $\cos(\theta) = 1$. This shows that anti-parallel configurations are much more favorable with two Arc-1-10 polymers than with two Arc-1-2 polymers.

\begin{figure}[ht!]
\includegraphics[scale = 0.55]{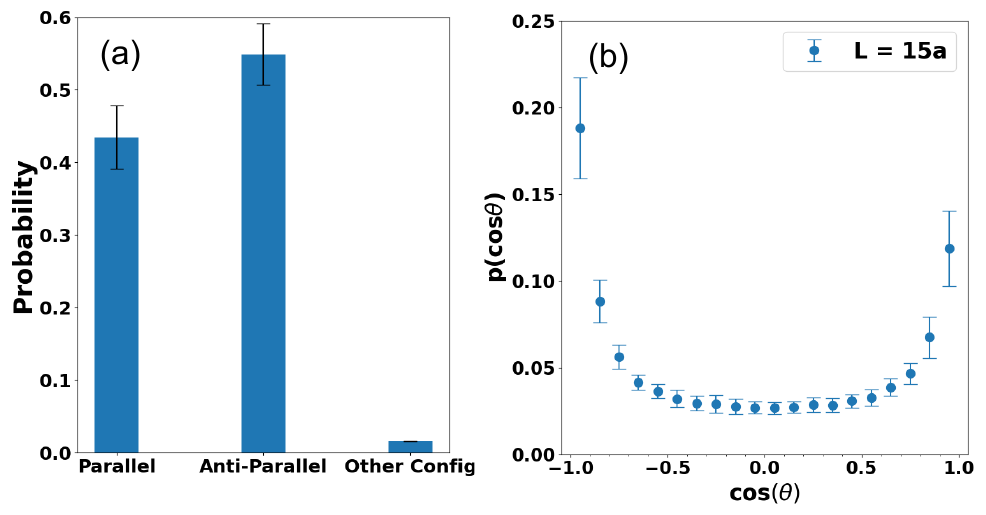}
\caption{(a) Probability of parallel and anti-parallel configurations with two Arc-1-10 polymers in a cylinder of length $15a$ using pbc along the $z$-axis. (b) As in (a) but now the distribution of $\cos(\theta)$ is shown. \label{fig7}}
\end{figure}

{\em Two Arc-1-10 polymers with PBC:} In the C4 configuration, the cluster of small loops point to the walls at the faces of the cylinder. Now we replace these walls by periodic boundary conditions (pbc) along the $z$ axis. As a consequence, the four distinct configurations in Fig.~\ref{fig3}d reduce to just two because the configurations C3 and C4 are equivalent with pbc. Similar to C1 and C2, their free energy must be equal. In the following, using pbc along the $z$ axis we denote C1/C2 as `parallel' and C3/c4 as `anti-parallel' configurations. The probabilities for the occurrence of these two types of configurations is shown in Fig.~\ref{fig7}a. We see again that there is a higher affinity for anti-parallel configurations for which the big loops of the two polymers are in close proximity and, due to pbc, the smaller subloops of the two polymers have to be adjacent to one another. To further substantiate this, we plot the probability distribution of $\cos(\theta)$ in Fig.~\ref{fig7}b. Again, we see that there is a clear preference for the anti-parallel configuration.

In a closed cylinder, the overlap of smaller subloops is highly unfavorable, as evidenced by the low probability of configuration C3 in Fig.~\ref{fig6}c for the Arc-1-10 polymer. However, with pbc we find that the smaller subloops face each other which is associated with a preference for an overlap of the big subloops. Furthermore, the free energy for overlaps of the two big subloops and small loops (in anti-parallel configurations) is less than the free energy of a pair of overlaps between the big subloop and small subloops in the parallel configurations (see below). As a next step, we consider more than two Arc-1-10 polymers confined in a cylinder with closed ends.

\begin{figure*}[ht!]
\includegraphics[scale = 1.5]{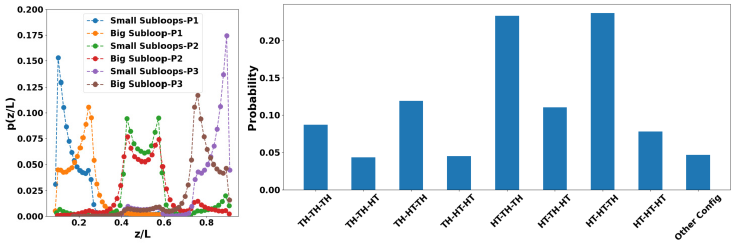}
\caption {(a) Probability distribution $p(z/L)$ of the COM of different subloops with three Arc-1-10 polymers. The three polymers are referred to as P1, P2 and P3. (b) Probabilities of observing different configurations of the three polymers along the $z$ axis. The configurations are labeled by the position of big subloop (B) and cluster of small subloops (S) of the three polymers along the $z$ axis from left to right. Each of the three polymer consists of $200$ monomers. The polymers are in a cylinder of length $22.5a$ and diameter $5a$. For this case with three Arc-1-10 polymers, we classify snapshots into eight most probable configurations (see text). Configurations with significant overlap of distinct polymers are counted as `Others'. \label{fig8}}
\end{figure*}

{\em Three Arc-1-10 polymers:} We first discuss the case of three polymers in a closed cylinder of length $L=22.5a$ (such that there is a cylinder length of $7.5a$ per polymer, as before). We measure the COM of the different subloops of each polymer and plot the probability distribution along the $z$-axis in Fig.~\ref{fig8}a. We observe six peaks corresponding to the six sets of subloops in the three polymers, indicating that the polymers (and their subloops) are well segregated and occupy three separate sections of the cylinder in most of the microstates. For the first polymer (P1) at the left of the cylinder, we see that there is a higher probability that the set of small subloops is located at the left face of the cylinder while the big subloop is preferentially away from the cylinder face. A similar behavior is, of course. found for the third polymer (P3) at the right end of the cylinder.

We also find that the peaks of $p(z/L)$ for the subloops correponding to P1 and P3 have a larger amplitude than those for P2 in the middle. The distribution for P2 is symmetric because the flipping of P2 in the middle of the cylinder will not cause any free energy difference. Furthermore, for P2, the distribution of the big subloop is slightly broader than that for the small loops. This implies that the big subloop of P2 haa more overlap with the respective big subloop of the neighboring polymer than the corresponding the small subloop. As a consequence, the peaks of $p(z/L)$ for the big subloop of P3 has a smaller amplitude than that for the corresponding small subloop. Note that the data is obtained from $40$ independent runs, and in some of the runs the P2-polymer has exchanged positions with the P1 or the P3 polymer. This is indicated by the non-zero probabilities to find the subloops of P2 near the faces of the cylinder.

\begin{figure*}[ht!]
\includegraphics[scale = 0.8]{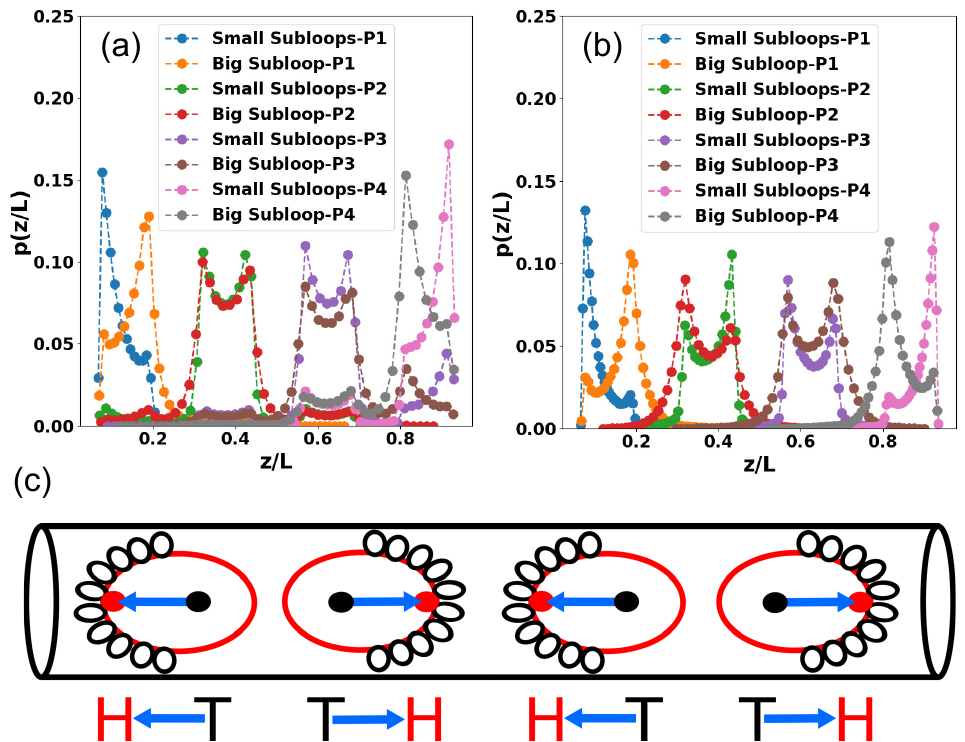}
\caption{(a) Probability density $p(z/L)$ along the long $z$ axis for the COM of four Arc-1-10 polymers ($N=200$) confined in a cylinder of length $L=30a$ and diameter $D_c=5a$. The four polymers are labelled P1, P2, P3 and P4. (b) As in (a) but now for a cylinder of length $L=70a$ and diameter $D_c=7a$, and each polymer with $N=500$ monomers. (c) Schematic of the most probable  (anti-parallel) configuration. \label{fig9}}
\end{figure*}

In the following, we refer to the small subloop as the head (H) and the big one as the tail (T) of the polymer. The orientation of the head is represented by a left or right arrow, i.e.~by $\leftarrow$ or $\rightarrow$, respectively. The walls (W) at the faces of the cylinder are represented by `$|$'. For example, a configuration where the tail of P2 points to the tail of P3 can be represented by $| \leftarrow \leftarrow \rightarrow |$. As in such a configuration, the central polymer (P2) overlaps with the big subloops on both sides, the big subloop of P2 has an equal probability of overlapping with the big subloop of the first (P1) and the third polymer (P3). To substantiate this, in Fig.~\ref{fig8}b we show the probabilities of all the eight possible configurations for a system of three polymers. We see that the probabilities of $| \leftarrow \leftarrow \rightarrow |$ and $|\leftarrow \rightarrow \rightarrow |$ are equal and, furthermore, these two configurations have the highest probability of all configurations. This is due to the occurrence of one overlap of big subloops and at the same time an overlap between a big and the small subloops of P2. All the other configurations involve at least one H-H overlap (as, e.g., in  $| \rightarrow \leftarrow \rightarrow |$) and thus have a lower probability (corresponding to a higher free energy). Configurations where both big subloops of P1 and P3 point toward the ends of the cylinder have an even lower probability and thus a higher free energy.

\begin{figure}[ht!]
\includegraphics[scale = 0.55]{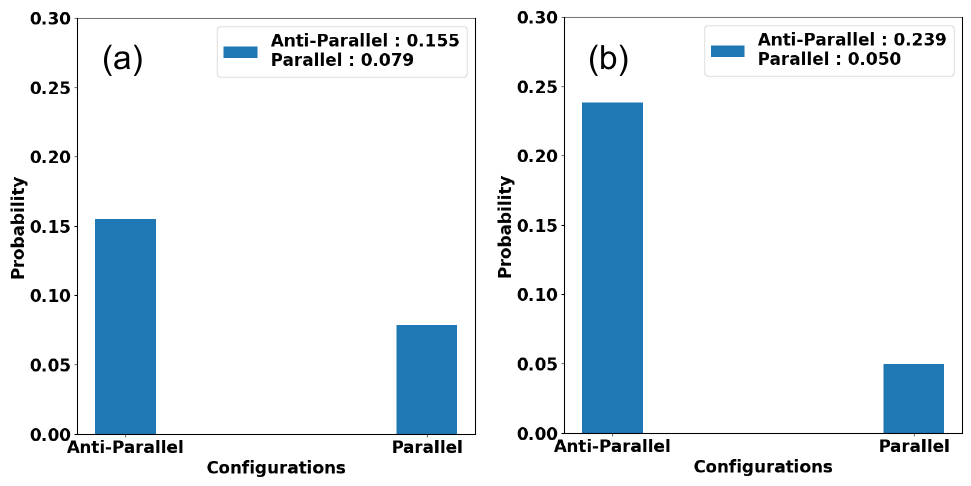}
\caption{(a) and (b) show the relative probabilities of `anti-parallel' and `parallel' configurations with $4$ Arc-1-10 polymers with $200$ and $500$ monomers, respectively. The cylinder diameters are the same as in the previous figure, i.e.~$5a$ and $7a$, respectively, but the length of the cylinders are $30a$ and $70a$, respectively, to accommodate $4$ polymers. \label{fig10}}
\end{figure}

{\em Four Arc-1-10 polymers:} Next, we confine four Arc-1-10 polymers with $200$ monomers each in a cylinder of length $L = 30a$, such that the monomer number density is maintained at the same value as before. We plot the spatial probability density distribution $p(z/L)$ of the COM of the subloops in Fig.~\ref{fig9}a. In Fig.~\ref{fig9}b, we consider configurations with $4$ polymers where each polymer has a total number of $500$ monomers (with $250$ monomers in the big subloop and 25 monomers in each of the 10 small subloops). As before, polymers remain segregated along the $z$ axis. However, for the smaller system $200$ monomers per polymer, we see that the third polymer has exchanged position with the fourth polymer in a relatively large number of microstates. Moreover, we also see that the first and  second polymers, as well as second and third polymers have exchanged positions with each other as reflected by non-zero probabilities in the corresponding cylinder segment(s). For the larger system (Fig.~\ref{fig9}b), we do not observe any exchange in the position of the polymers. Moreover, the difference in probabilities near the cylinder ends for small and big subloops is significantly larger for the system with $500$ monomers per polymer, as compared to that with $200$ monomers per polymer. Both this observations indicate that the free energy barrier for flipping is higher for the longer polymer.  

Upon analyzing the relative probabilities of the arrangement of subloops, we find that the polymers preferably arrange such that the big subloops overlap while at the same time there is a minimal overlap of small subloops. The most probable configuration for four polymers is depicted in Fig.~\ref{fig9}c. We refer to this arrangement as the `anti-parallel' configuration (note that it contains one H-H overlap). The `parallel' configuration is W-H-T-H-T-H-T-H-T-W or W-T-H-T-H-T-H-T-H-W. It contains no overlaps between big subloops.

In Figs.~\ref{fig10}(a) and (b), we plot the probabilities for the occurrence of parallel and anti-parallel configurations for systems with 200 and 500 monomers per polymer, respectively. The probability of the `anti-parallel' configuration is significantly higher than that of the `parallel' configuration. All the other cases are not plotted here as there are 16 different configurations with four polymers. We see that for the large system, the probability for the `anti-parallel' configuration is five times larger than that for the`parallel' configuration. It is important to note that there are two T-T overlaps in the anti-parallel configuration but also one H-H overlap (Fig.~\ref{fig9}c). In the `parallel' configuration there are four $H-T$ overlaps, but there is no $H-H$ overlap. Thus, we can conclude that two T-T plus one H-H contact has a lower free energy than any other configuration with four polymers. A better understanding of these observations can be obtained via the analysis of the free-energy landscape of the different configurations. We carry out this analysis in the next two sections.

\begin{figure*}[ht!]
\includegraphics[scale = 0.9]{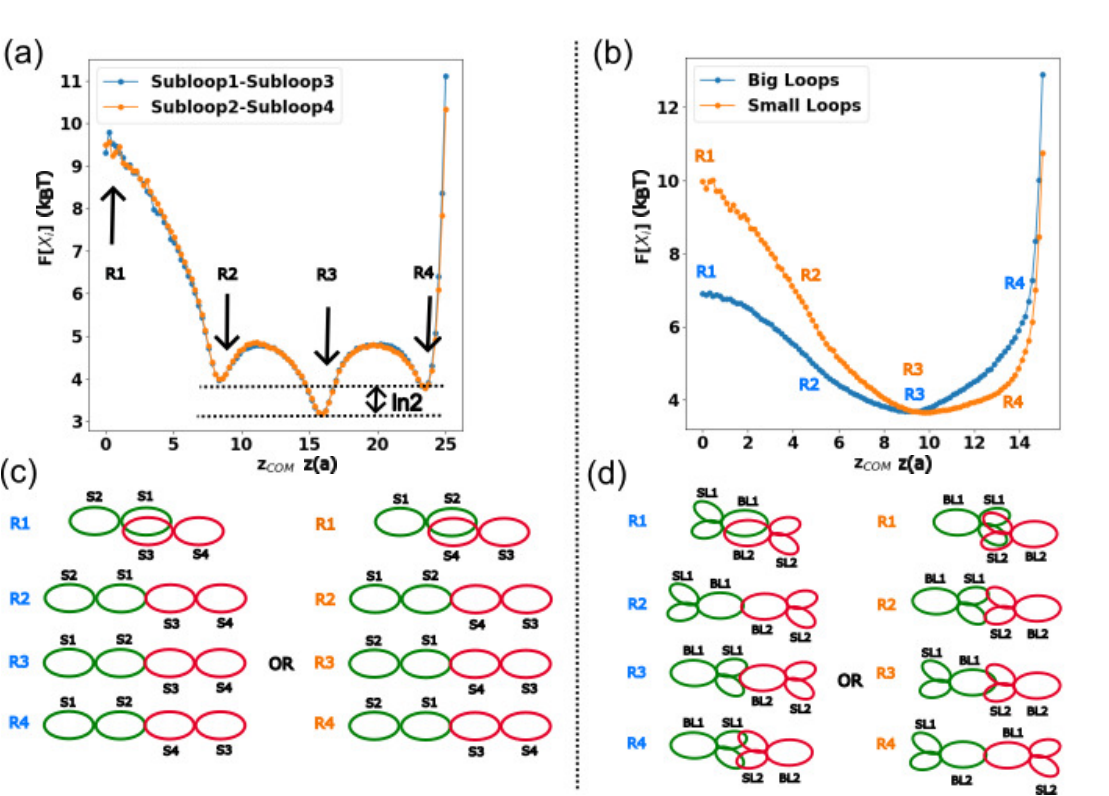}

\caption{ (a) Free energy $F[X_i]$ in units of $k_B T$ for two `rotated-8' polymers, each with $N = 200$ monomers, in a cylinder of length $25a$. The reaction coordinate $X_i$ is $X_i= z_{com}$, corresponding to the distance between the CoMs of particular loops (see text). Four different regions of $F[X_i]$ are marked as R1, R2, R3 and R4, and the corresponding arrangement of the loops at these values of $X_i$ are shown in the schematic below (a), each of which is different for small loops (SL) and big Loops (BL). (b) Free energy $F[X_i]$ for two Arc-1-2 polymers, each with $200$ monomers. As in (a), the values of $F[X_i]$ for the four different polymer configurations are marked in the figure and the corresponding configurations are shown in the schematic below. \label{fig12}}
\end{figure*}

\begin{figure*}[ht!]
\includegraphics[scale = 0.9]{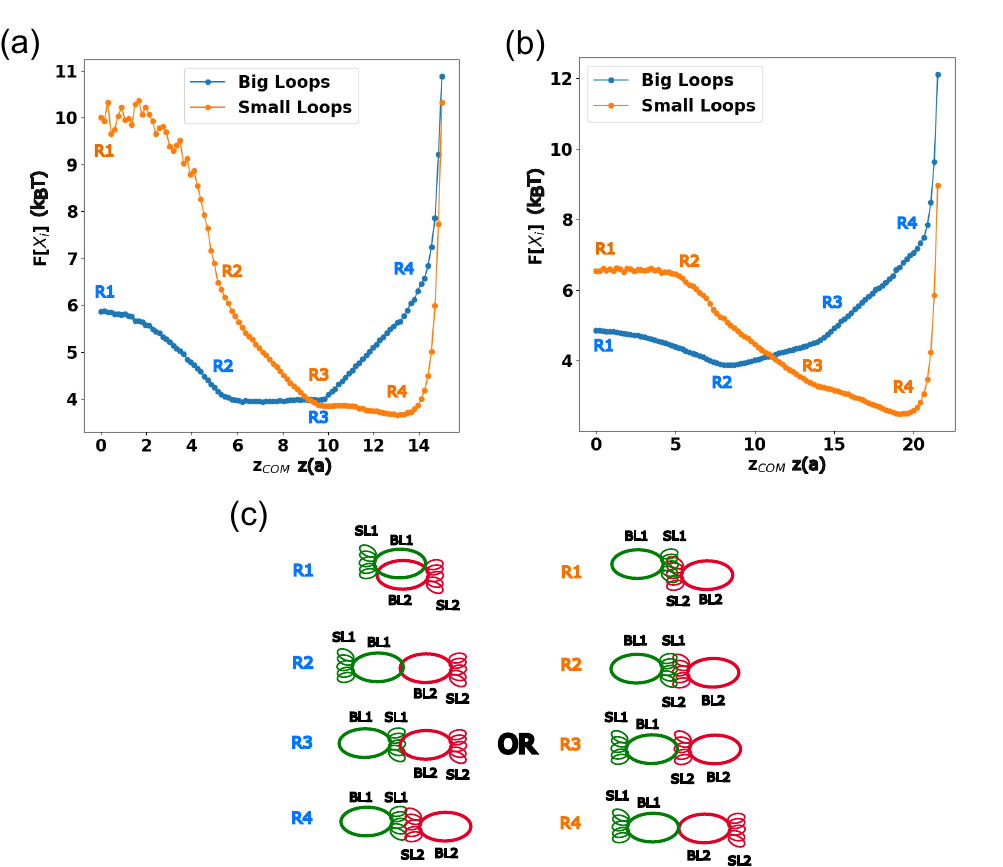}
\caption{(a) Free energy $F[X_i]$ in units of $k_BT$ for two Arc-1-10 polymers, each with $N = 200$ monomers, confined in a cylinder of length $L=15a$ and diameter $D=5a$. (b) Free energy $F[X_i]$ for a pair of polymers, each with $N=500$ monomers, confined in a cylinder of length $L=21a$ and diameter $D=7a$. The reaction coordinate $X_i$ is the distance between the COMs of two clusters of small loops (SL) and COMs of big loops (BL) with polymer index $1$ and $2$. These are marked in the legend. (c) A schematic of polymer arrangements in four different scenarios, marked as R1, R2, R3 and R4. \label{fig13}}
\end{figure*}

\section{Free energy differences: C1-C2-C3-C4} 
We have discussed the probabilities for the occurrence of configurations C1, C2, C3, C4 (see Fig.~\ref{fig3}d). To explore the free energy landscape of the two confined polymers for the three different topologies, i.e.~`rotated-8', Arc-1-2 and Arc-1-10, we need to define appropriate reaction coordinates. For the asymmetric topologies, namely Arc-1-2 and Arc-1-10 with big loops and cluster of small loops, we use two reaction coordinates to fully describe the different arrangements of polymers with respect to each other: (i) the distance between the COMs of the big loops of the two polymers, $X_1$, and (ii) the distance between the COMs of the cluster of small loops of each polymer, $X_2$. For the `rotated-8' topology, the two loops of each polymer are of equal size (cf.~Fig.~\ref{fig1}b). In this case, we define $X_1$ as the distance between the COMs of subloop-1 and subloop-3 and $X_2$ as the distance between the COMs of subloop-2 and subloop-4.
   
In the following, we estimate free energy differences between characteristic configurations for the different polymer topologies. As before, averages are taken over $40$ independent runs, where in each run with a total length of $10^8$ iterations we collected data every $10^4$ iterations and thus used at least $N_T = 4 \times 10^5$ statistically independent configurations for the analysis. We calculate the probability $p[X_i]$ that a particular value of $X_i$ is observed in $N_i$ configurations out of $N_T$ configurations ($i=1,2$). We then determine the free energy $F[X_i]$ as a function of $X_i$ as given by
\begin{equation}
F[X_i] = -k_BT \, \ln(p[X_i]) + C \, ,
\end{equation}
where  $C$ is a constant that is independent of $X_i$, and thereby irrelevant since we are only interested in free energy differences. Therefore, we set $C=0$. Figures \ref{fig12} and \ref{fig13} show $F[X_i]$ for the polymer topologies `rotated-8', Arc-1-2, and Arc-1-10.
We now discuss the free energy landscape of the three topologies one by one. 

{\em `Rotated-8'}: Due to the symmetry of the `rotated-8' topology, the free energies with respect to  $X_1$ and $X_2$ are identical (cf.~Fig.~\ref{fig12}(a)). Therefore, we only discuss the results for $F[X_1]$. In Fig.~\ref{fig12}a, a value of $z_{COM} = X_1 \approx 0$ corresponds to a complete overlap between subloop-1 and subloop-3, leading to a high value of free energy. The configurations for $X_1 \approx 0$ corresponds to the schematic denoted as R1 in Fig.~\ref{fig12}(c). An increase of $X_1 \equiv z_{COM}$ corresponds to a decrease of the overlap between subloop-1 and subloop-3 and a decrease of the free energy $F[X_1]$ until there is no overlap between the two loops, reaching the configuration R2. As $z_{COM}$ ($=X_1$) increases further, an overlap between the two subloops of the same polymer occurs. Thereby, $F[X_1]$ starts to increase as the system explores configurations with an increase in overlap of the two subloops. The overlap of internal loops of a polymer increases till  one of the polymers gets fully flipped. Thereafter, $F[X_1]$ shows a global minimum at $X_1$ which is associated with the configurations R3. This global minimum is a consequence of $X_1$ having the same value irrespective of whether polymer P1 or P2 flips. Hence, there is a contribution from two equivalent sets of configurations, corresponding to R3. 

A further increase in $X_1$ corresponds to configurations with overlap of internal subloops of the un-flipped polymer, with a corresponding increase in $F[X_1]$. A flip in the second polymer occurs for higher values of $X_1$. After the flip, subloop-1 and subloop-3 are both proximal to the cylinder poles in the R4 configuration and again corresponds to a minina in $F[X_1]$. Further increase in $X_1$ corresponds to a strong overlap of subloop-1 or subloop-3 with the wall at the poles. The depth of the global minimum can be verified as $\ln{2} k_BT \approx 0.7 k_BT$ less than the values of $F[X_i]$ at the two symmetric minima.

{\em Arc-1-2 polymers:} In Arc-1-2 polymers, the presence of one big loop cross-linked to two conjoined small loops leads to an asymmetry of the polymer topology. Now, the free energy $F[X_i]$ depends on the choice of $X_i$, i.e., whether the distance between big loops (BLs) or small loops (SLs) is chosen as a reaction coordinate. The configurations with overlap of SL1 and SL2 is much less probable than that where BL1 and BL2 overlap. This is reflected by the fact that the free energy difference with respect to $X_1$ is different from that with respect to $X_2$, especially for low values of the chosen reaction coordinate. represented by the configuration R1 in the schematics.  

We first discuss the case $z_{COM}= X_1$, with $X_1$ the distance between BLs of two polymers. An increase in $X_1$ corresponds to a reduction of the overlap between BLs and, as a result, configuration R2 is obtained. In the set of R2 (blue) configurations, the BLs occupy the center of the cylinder without overlap. Thus, R2 is similar to the C4 configuration discussed before. However, when $z_{COM}=X_2$, R2 configuration corresponds to C3 and then $F[X_2] > F[X_1]$ holds. With a further increase of $X_1$ (or $X_2$), one of  the polymers goes through a flip. As a consequence, one reaches R3 configuration corresponding to the global minima in $F[X_1]$ (and $F[X_2]$). Thereby, a global minimum is seen irrespective of whether one chooses $X_i$ to be $X_1$ or $X_2$. The minimum corresponds to the two configurations R3 (blue) and R3 (orange), both of which are equivalent and degenerate with respect to $F[X_i]$. The contributions from both R3 (blue) and R3 (orange) lead to the global minimum in $F[X_i]$ at $X_i \approx 10\,a$.   
For large values of $X_1$ ($X_2$), there is a strong increase of $F[X_1]$ ($F[X_2]$) and one observes configuration R4 (blue) [R4 (orange)], wThe R4 (orange) being more probable than R4 (blue), such that around R4 $F_[X_1]>F[X_2]$ holds. A further increase in $X_1$ (and $X_2$) leads to a decrease of the distance between the poles and the monomers of BLs (and SLs), which in turn increases $F[X_1]$ (and $F[X_2]$) due to an increasing repulsive interaction.

{\em Arc-1-10 polymers:} After the analysis of the free energy landscape of the relatively simple cases with two `rotated-8' polymers and two Arc-1-2 polymers, we now analyze the case of two Arc-1-10 polymers. We consider two system sizes with $N=200$ and $N=500$ monomers per polymer. The corresponding results are respectively shown in Figs.~\ref{fig12}(a) and (b). 

The difference in the behavior of $F[X_1]$ and $F[X_2]$ between the Arc-1-10 and Arc-1-2 systems can be understood in terms of the underlying probability distributions: for Arc-1-10 polymers, the combined probabilities of C1 and C2 can be comparable to or smaller than that of C4. This effect becomes more pronounced for longer chains ($N=500$). So the global minimum of $F[X_1]$ ($F[X_2]$) is now at values of $X_1$ ($X_2$) that correspond to configuration C4. As before, configurations C1 and C2 cannot be distinguished via the reaction coordinates $X_1$ and $X_2$. The schematics in Fig.~\ref{fig13}(c) illustrate the corresponding configurations for the entire range of values of $X_1$ and $X_2$.

The R1 (blue) and the R1 (orange) configuration correspond to $X_1 \approx 0$ and $X_2 \approx 0$, respectively (Fig.~\ref{fig13}a). Here, the difference between $F[X_2]$ and $F[X_1]$ is much higher compared to the case of Arc-1-2 polymers. For $X_1 \approx 5a$, one obtains R2 (blue) where the overlap between BLs is minimal. For the larger system with $N=500$, this configuration corresponds to a global minimum of $F[X_1]$ (Fig.~\ref{fig13}b). At $X_1 \approx 10a$,  the R3 (blue) configuration is located where both polymers are orientated parallel to each other with only one of the SLs closer to the pole. As clarified previously, $F[X_1]$ at $X_1 \approx 10a$ has contributions from both C1 and C2. For the smaller system with $N=200$, the free energy $F[X_1]$ at $X_1 = 5a$ and $10a$ is similar, while for $N=500$, $F[X_1]$ corresponding to R3 (blue) is higher than $F[X_1]$ corresponding to R2 (blue). Higher values of $X_1$ correspond to R4 (blue),

We now discuss $F[X_2]$ shown in Figs.~\ref{fig13}(a,b). For small values of $X_2$ (i.e.~$X_2 \to 0$), there is an overlap of the SLs of the two polymers. This overlap gradually decreases for values of $X_2$ greater than $6a$ (R2: orange) in Fig.~\ref{fig13}a and in $5a$ to $10a$ in Fig.~\ref{fig13}b. For $N=200$ Values of $X_2 \approx 10a$ in Fig.~\ref{fig13}a correspond to the R3 (orange) configuration while for $N=500$ $X_2$ is around $14\,a$ at the R3 (orange) configuration. The global minimum in $F[X_2]$ at R4 (orange) is more pronounced for the $N=500$ system than for the $N=200$ system.  

\begin{figure*}[ht]
\includegraphics[scale = 0.8]{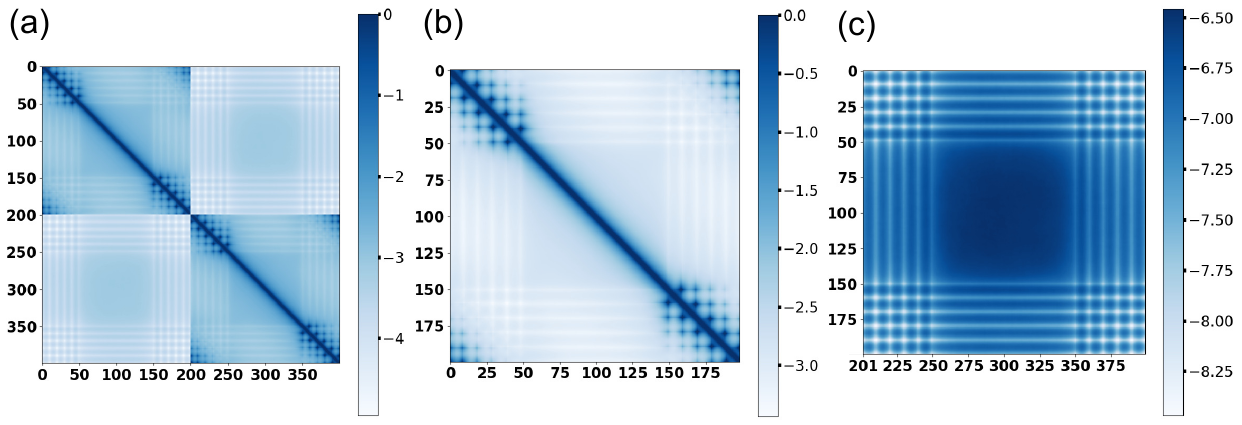}
\caption{(a) Contact probability map of two Arc-1-10 polymers with $200$ monomers in each polymers confined in a cylinder of length $15a$ and with periodic boundary conditions along the $z$-axis. The colors in the color map correspond to contact probabilities on a logarithmic scale (base 10). (b) Zoomed version of the quarter of the contact map to better decipher the intra-polymer contacts. (c) Zoomed version of the quarter of the contact map which shows inter-polymer contacts, i.e.~contacts between polymers P1 and P2. Different range of color-scales are used in the color maps of panels (a), (b), and (c), to obtain an optimal visual contrast for the different cases. \label{fig14}}
\end{figure*}

\section{Contact Map of two confined Arc-1-10 polymers}
Since the advent of HiC maps \cite{LiebermanAiden2009}, there have been many  studies to 
interpret the contact maps of chromosomes to get an understanding of the 3d-organization 
of chromosomes \cite{Haddad2017,Agarwal2018, Agarwal2019,Agarwal2019_2}. 
The contact map represents the probability that two chromosomal segments are in contact. Hi-C maps, obtained from large ensembles of cells at various stages of the cell cycle, thus capture an average over diverse chromatin conformations. Differences in contact patterns arise not only from stochastic fluctuations in polymer configurations but also from the action of specific proteins that bridge distinct DNA segments, such as promoters and enhancers, to regulate gene expression. Consequently, the formation or disruption of promoter–enhancer contacts alters the overall contact probabilities between chromosomal segments. Hi-C maps therefore reflect ensemble-averaged contact probabilities encompassing all such molecular interactions.

In contrast to the transient contacts observed in cellular environments, we study polymers with fixed cross-links between specific segments. From our simulations, we obtain the contact probability between any pair of polymer segments. Using the contact map, we can quantify how different polymer segments 
interact in the presence of permanent cross-links. 
The contact map provides insight into the organization of monomers within clusters of small loops and furthermore reveals how exposed the cross-linked sites are to contacts with other monomers, either within the same polymer or from neighboring chains. Such analyses also serve as a basis for understanding more complex systems with transient cross-links or multiple interacting polymers.

Figure~\ref{fig14}a shows the color map of the contact matrix for 
two Arc-1-10 polymers with $200$ monomers each, confined in a cylinder.
The matrix element $M_{ij}$  represents the  probability that  a monomer $i$ is in contact 
with a monomer $j$. Two monomers  are said to be in contact if the distance $d_{ij}$ 
between the two monomers $i$ and $j$ satisfies the condition
\begin{equation}
d_{ij} \le 1.5a \, .
\label{eq_dij}
\end{equation}
Here, the indices $i, j = 1, \dots, 200$ correspond to the monomers of the first polymer P1 and $i, j = 201, \dots, 400$ to the monomers of the second polymer P2. The indices $51-100-150$ ($251-300-350$) belong to the big subloop of polymer P1 (polymer P2), while the other indices correspond to monomers in small subloops. From the definition 
in Eq.~(\ref{eq_dij}), the diagonal elements $M_{ii}$ and the nearest-neighbor 
elements $M_{i,i\pm1}$ are unity for all $i$. In the following, we analyze the color maps 
of contact probabilities, where the color scale is logarithmic and represents the magnitude 
of the contact probability between monomer pairs.

Figure~\ref{fig14}a shows the contact map of two Arc-1-10 polymers, which exhibits two dark 
and two light-colored blocks. The dark blocks correspond to intra-polymer contacts, while 
the lighter ones represent inter-polymer contacts. The presence of distinct blocks indicates 
that the polymers remain largely segregated. To gain insight into the organization of the 
smaller loops, we magnify one intra-polymer block (Fig.~\ref{fig14}b) and one inter-polymer 
block (Fig.~\ref{fig14}c). The color scales in Figs.~\ref{fig14}b and \ref{fig14}c differ 
from that in Fig.~\ref{fig14}a to enhance visual contrast.
 
In the intra-polymer contact map, apart from the region along the diagonal, we observe a central 
dark square and a grid-like pattern at the top-left and bottom-right corners (Fig.~\ref{fig14}b). 
The central dark square corresponds to contacts among monomers forming the large loop ($51$–$100$–$149$), while the grid-like pattern at the four corners arises from contacts among monomers within clusters of small subloops. Monomers connected by springs of length $a$ to create the ten cross-links remain in close spatial proximity, exhibiting a contact probability of unity and forming a regular pattern of dark spots. In contrast, monomers distant from the cross-links are not always in contact, as defined by our contact criterion. 

Now we discuss the top, bottom, left and right striped regions surrounding the central square.
Lighter lines correspond to lower contact probabilities between monomers of the big loop (BL) and those of the small subloops (SSLs), indicating that SSL monomers are relatively shielded from the BL. 
White stripes at the top of the central dark square correspond to monomers forming the cross-links; 
these  generate also the dark grid points in the corner regions of Fig.~\ref{fig14}b. 
Consequently, non–cross-linked monomers of the SSLs exhibit higher contact probabilities with BL monomers 
than do the cross-linked monomers, suggesting a rosette-like arrangement of small loops, 
with the cross-links buried near the rosette’s center.

In Fig.~\ref{fig14}b, the bottom-left (or equivalently, the top-right) corner of the intra-polymer contact map shows contacts between the five small subloops on the left arm and the five on the right arm. The saturated dark regions correspond to monomers forming the cross-links, while parts of these corner grids display very low contact probabilities. This indicates that the subloops containing monomers $150$–$160$–$170$ (on the left arm) have relatively low contact probabilities with those comprising monomers $30$–$40$–$50$ (on the right arm). Hence, even the small subloops tend to repel each other and remain spatially well separated.

Next, we examine the inter-polymer contact map shown in Fig.~\ref{fig14}c. A central dark square corresponds to contacts between the big loops (BLs) of the two polymers. This is consistent with Fig.~\ref{fig4}b,c, where the BLs are shown to occupy the central region of the confining cylinder, leading to a higher frequency of contact. On closer inspection, a grid-like pattern appears in the four corners of the central dark block, representing contacts between the sets of small subloops (SSLs) from the two polymers. The grid structure indicates that the inner monomers of the loops are shielded, and that SSL–SSL contacts occur primarily between monomers located away from the cross-links. The relatively lighter stripes along the top, bottom, left, and right of the central dark square indicate lower contact probabilities between the cross-linked monomers forming the SSLs and the monomers of the BL. Thus, contacts between BL monomers arise predominantly from regions distant from the cross-links of the SSLs.

\section{Discussions}

In summary, we add topological modifications to a bead spring model of ring polymer by
adding suitable  cross-links  such that there is a rosette of  smaller loops attached to 
a bigger loop. A pair (or more) of such  polymers are confined within a cylinder. 
The monomers of the polymer interact with each other with purely
repulsive,  self-avoiding interactions. The polymers not only remain segregated due to 
entropic reasons, but also arrange themselves in a manner such that the rosette of loops 
try to avoid each other and face the poles of the cylinder. If one draws a vector from  the 
center of mass of the big loop to the center of mass of the rosette of loops, one can 
claim that the two vectors   (say $\vec{V_1}$ and $\vec{V_2}$) representing the two 
polymers have a preference for anti-parallel alignment.
However, it is pertinent to point out that preference for such a configuration is entropically driven, 
as we have implemented  only repulsive interaction between any pair of monomers. If the rosette of loops
in each polymer face the poles, the   big loops from the two polymers can overlap at the center of the 
cylinder and explore more configurations to  increase entropy. In contrast, the rosette of loops behave
more like a cluster of small soft spheres and try  to minimize overlap.

There are essential differences between the alignment of polymers along the long axis of 
a cylinder in our model, and  the anti-ferromagnetic interaction (Hamiltonian) of spins 
in the Ising model.  Two Ising spins can either be parallel or anti-parallel to each other 
with energies $-J$ or $J$, respectively. 
In contrast, in our case of two polymers  in the cylinder, we can have  four different alignments of 
the two vectors, out of which the anti-parallel  alignments of the polymers are not equivalent. 
The system will have the same free energy in the two cases when $\vec{V_1}$ and $\vec{V_2}$ are parallel 
to each other. But the two different antiparallel configurations (vectors facing either 
towards or away from each other) correspond to two different values of free energy. 
Thus, the system prefers to be in one of the three different states corresponding to 
free energy minima. A similar three-state system  has been described to model contact 
inhibition  locomotion on confined cellular organization in Ref.~\cite{Muhuri}.
However, the model introduced in this reference describes three distinct  
values of  {\em energies} possible for 
four configurations of vectors, whereas the different configurations in our system arise  due to 
{\em entropic} considerations. If we implement periodic boundary conditions in our system, 
the three-level system reduces to a two-level system since the anti-parallel  configurations become equivalent as well.

Tuning effective interactions between polymer segments by designing internal loops is thus a useful 
additional handle to achieve a degree of polymer organization in confinement, which in turn is agnostic 
to the detailed energetic interactions between polymer segments. Although, in the current work we have 
looked at clusters of loops packed within a cylinder, it is also relevant to investigate organization 
of multiple topologically modified polymers within other confining geometries  such as  spheres. 
This is investigated in Ref.~\cite{Kingkini}. 
The choice of confining geometries have  been motivated by
the confinement of chromosomes in the bacterial cell, as well as the shape of the nucleus in 
eukaryotic cells.  In fact,  when investigating longer polymers, the size of the confining cylinder  
(and  the sphere) has been chosen to keep  the monomer density and volume fraction to be fixed at 
$0.2$, as the chromosome volume fraction in cells is estimated to vary between $0.1$ and $0.2$. 
In polymer physics investigation it is natural to keep the ratio of radius of gyration(s) 
of the polymer(s) and the diameter $D$  fixed when comparing different scenarios.

While this work was initiated with the aim of understanding the orientational alignment of 
ToMo polymers,  our investigation also provides insights into the effective entropic 
interactions between clusters of small  loops from different polymers, as well as between 
large and small loops and between the large loops themselves. 
These interactions offer a physical basis for the mechanisms governing chromosome organization 
and dynamics.  Notably, we are adapting the Arc-1-10 polymer topology and the emergent 
organizational features identified  here to study chromosome organization in the bacterial cell 
of {\em C.~crescentus} \cite{Badrinarayanan2015}.  The Arc-1-2 topology was used to explain 
chromosome organization in {\em E.coli} \cite{dna1,dna2,fast_growth}.  
Furthermore, the  free-energy landscape associated with polymer 
overlap is being employed to explore the segregation kinetics  of ToMo polymers, which in turn will aid 
in estimating chromosome segregation time  scales in cylindrical  bacterial cells.

However, the principles of entropically driven polymer organization outlined here remain relevant 
for synthetic polymers as well. In a melt of block copolymers, it is possible to design various 
micro-structures by inducing microphase separation by tuning the interaction and relative lengths
of the different segments of the  block copolymer. It would be interesting to investigate if 
similar organization at higher length scales can be achieved with topologically modified polymers. 

\section{Author Contributions}
S.B. implemented the model and performed the calculations and data analysis. The initial research plan was conceived by D.M. and A.C. S.B. and A.C. discussed the project extensively with J.H., who supervised the free-energy calculations. S.B., D.M., J.H., and A.C. jointly wrote the manuscript.
\section{Acknowledgements}
Authors acknowledge useful discussions with Arieh Zaritsky, Conrad Woldringh, Peter Virnau, and Sathish Sukumaran. 
AC, with DST-SERB (IN) identification No. SQUID-1973-AC-4067, acknowledges funding by DST-SERB (IN) 
project CRG/2021/007824, funding by Biosantexc research and mobility programs funded by ENS-France
and opportunity  to host 
Prof. Jie Xiao using funds from the Fulbright-Nehru specialist program from USIEF
A.C also acknowledges discussions in meetings organized by ICTS, Bangalore, India and use of the computing facilities 
by PARAM-BRAHMA. 

\bibliographystyle{unsrt}
\bibliography{ref.bib} 

\end{document}